\documentclass[aps,prl,twocolumn]{revtex4-1}
\usepackage{amssymb}
\usepackage{amsmath,bm}
\usepackage{graphicx,color}
\usepackage{bbm}
\usepackage{multirow}
\newcommand{\B}[1]{{\bm{#1}}}

\usepackage{hyperref}

\begin{document}
\title{Thermodynamic Equivalence of Cyclic Shear and Deep Cooling in Glass-Formers}
\author{Bhanu Prasad Bhowmik$^1$, Valery Iylin$^1$ and Itamar Procaccia$^{1,2}$}
\affiliation{ $^1$ Dept. of Chemical Physics, The Weizmann Institute of Science, Rehovot 76100, Israel\\ $^2$Center for OPTical IMagery Analysis and Learning, Northwestern Polytechnical University, Xi'an, 710072 China}
\date{\today}

\begin{abstract}
The extreme slowing down associated with glass formation in experiments and in simulations results in serious difficulties to prepare deeply
quenched, well annealed, glassy material. Recently, methods to achieve such deep quenching were proposed, including vapor deposition on the experimental side and
``Swap Monte Carlo" and oscillatory shearing on the simulation side. The relation between the resulting glasses under different protocols remains
unclear. Here we show that oscillatory shear and ``Swap Monte Carlo" result in thermodynamically equivalent glasses sharing the same
statistical mechanics and similar mechanical responses under external strain. 
\end{abstract}
\maketitle

Typical glass formers exhibit a dramatic slowing down in their dynamics when the temperature is reduced \cite{09Cav}. A rapid
quench to low temperatures results in a poorly annealed glass which may take exceedingly long time to age and evolve towards deeper
energy states \cite{95VKB,13JBP}. The study of well annealed glasses using numerical simulations is particularly sensitive to this fundamental limitation,
it simply takes too long to cool down an equilibrated model glass former of any useful size (number of particles) to temperatures that are comparable
to what can be achieved in experiments. Some remedies to this fundamental difficulty were proposed recently. On the experimental side
the creation of ultra-stable glasses was announced using vapor deposition \cite{07KSETY,13LEP}. For theoretical work this method has disadvantages in forming
anisotropic samples \cite{15LLWREP}. Two other methods were advanced for attaining similarly well quenched glasses in numerical simulations. One is
``Swap Monte Carlo" \cite{01GP} and the other is cyclic shearing \cite{13FFS}. The former approach adds to the usual acceptance or rejection of Monte Carlo steps
also the swap of two different particles. Such a swap would take enormously long times in standard Monte Carlo or in molecular dynamics,
since it will require a lengthy series of small moves. The main success of this method is the ability to create low temperature
glasses {\em without falling out of equilibrium}. Cyclic shearing starts first with a fast quench to a wanted low temperature, and
then subjecting the resulting poorly annealed glass to oscillatory shear strain with a chosen maximal strain such that $\gamma^{xy}(t) =
\gamma_{\rm max}^{xy} \cos{\omega t}$. It was claimed that such a protocol takes the poorly annealed glass through a series of states
that end up with well annealed glasses of competitive stability to those obtained via the Swap Monte Carlo protocol \cite{17LPS,18DPS}. 

In fact, it appears that these claims suffer from a lack of good criteria to compare the configurations attained by the competing
protocols. In particular is was not clarified whether with cyclic shear the systems do not fall out of equilibrium.
In this Letter we propose such criteria, allowing us to state that Swap Monte Carlo and oscillatory shearing can indeed
be shown to end up with equivalent thermodynamic states. Importantly, they both do not fall out of equilibrium. 
A key idea for achieving this capability is to execute both protocols in NPT
rather than NVT ensembles, allowing the density of the glass former to freely respond to the configuration changes imposed by
both protocols. In addition, in cyclic shear one restricts $\gamma_{\rm max}^{xy}$ to be smaller than the yield strain of the material.
We show below that these simple changes in conditions allow us to demonstrate that the different protocols
result in configurations that satisfy in the mean the same equations of state,  with identical probability distribution function (pdf)
of energy fluctuations about the mean state. In addition, the mechanical response to external shear strain is also shown to follow
the same stress vs. strain curves.  

To demonstrate these results we first simulate two types of systems, a standard binary glass former of $N$ point particles with 
inverse power law potential and a poly-dispersed glass former of $N$ point particles with the same potential. Below we comment
also on models with generic potentials like Lennard-Jones. The first system partially
crystallizes at sufficiently low temperatures \cite{15GKPP}, but the second does not \cite{19BFFSS}. Nevertheless both examples demonstrate
the main tenets of this Letter. The point particles are interacting via two-body interactions $\Phi(r_{ij})$ that depend
on the distance $r_{ij}\equiv |\B r_i -\B r_j|$ where $\B r_i$ is the position of the $i$th particle. In the first
example the system is made of 50-50 mixture of particles ``a" and particles ``b" with the potential reading 
\begin{equation}
\Phi_{ab}(r_{ij}) = \epsilon \left(\frac{\sigma_{ab}}{r_{ij}}\right)^{12 } \ , \quad \text{Example I} \ .
\label{defphi}
\end{equation} 
Here $\sigma_a=1.0$ and $\sigma_b =1.4$, $\sigma_{ab} = (\sigma_a+\sigma_b)/2$.

The second example is composed of point particles whose range of interaction $\sigma_i$ is taken from 
a distribution $P(\sigma)$ that reads
\begin{equation}
P(\sigma) = \frac{A}{\sigma^3}\ , \quad \sigma_{\rm min} \le \sigma \le \sigma_{\rm max}\ , 
\end{equation}
where $A$ is a normalization constant. Here $\sigma_{\rm min}=0.73$ and $\sigma_{\rm max}=1.62$ \cite{19BFFSS}.
In this model two particles $i$ and $j$ are interacting via the same power law Eq.~(\ref{defphi}) but with
\begin{equation}
\sigma_{ij} =\frac{\sigma_i+\sigma_j}{2} \left(1-0.2 |\sigma_i-\sigma_j|\right)\ , \quad \text{Example II} \ .
\end{equation}
The unit of length is $\sigma_a=1$ in the first example and $$\langle \sigma\rangle =\int_{\sigma_{\rm min}}^{\sigma_{\rm max}} d\sigma P(\sigma) \sigma =1$$ in the second example. The energy scale
is $\epsilon=1$. 

\begin{figure}
	\includegraphics[width=0.38\textwidth]{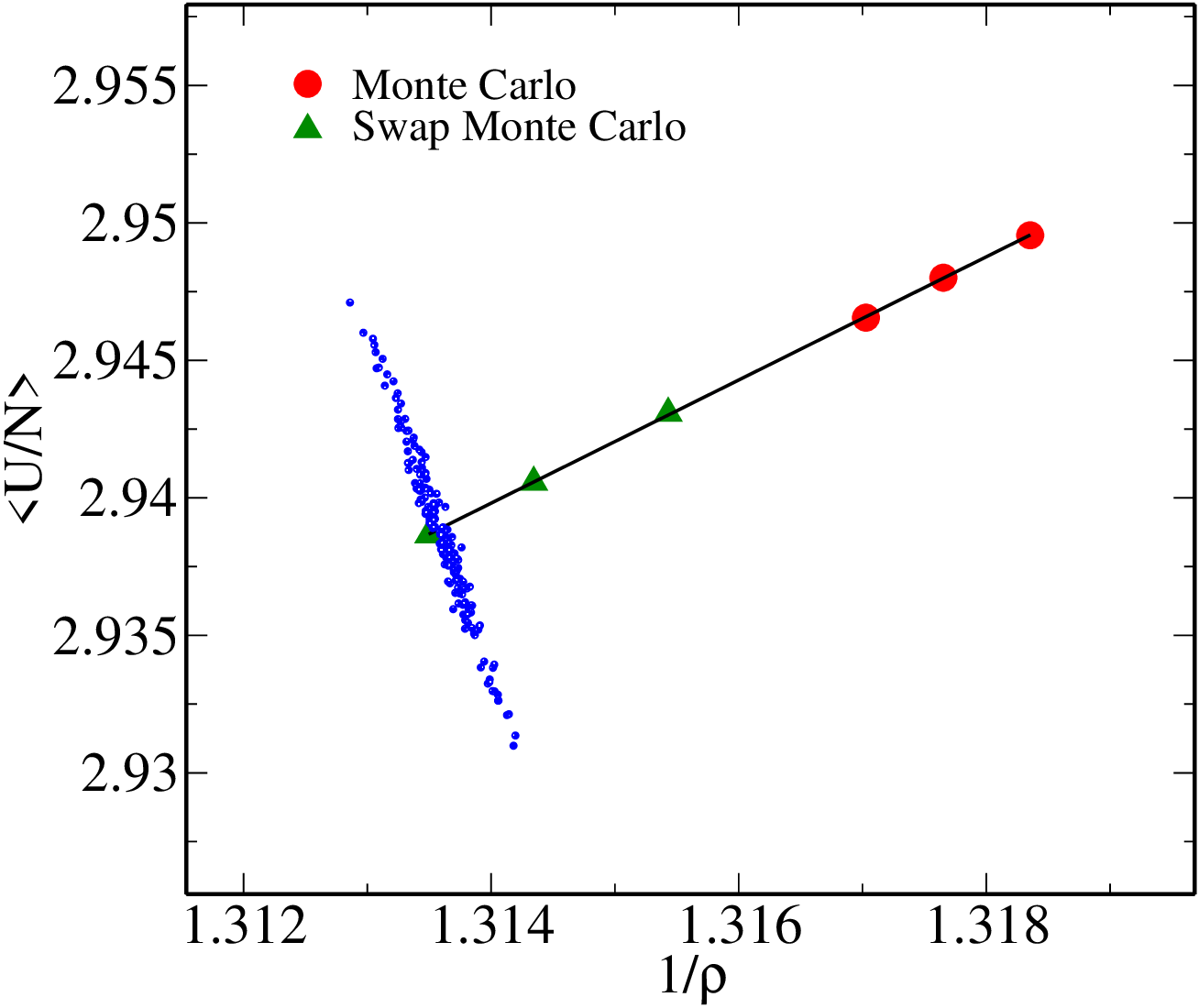}
	\caption{Average energy per particle in a binary system with $N=1024$ as a function of the inverse density
		after cooling at six different rates, three using regular Monte Carlo and three Swap Monte Carlo. 
		The best fit has a slope of 2.24$\pm 0.01$ in perfect agreement with the equation of state Eq.~(\ref{eqstate}).
	Note that Eq.~(\ref{eqstate}) is valid only on the average. To underline this, typical fluctuations in energy and density
leading to the lowest average energy in this graph are shown in blue. }
	\label{coolbin}
	\end{figure}

The preparation protocol is the same for both examples. All simulations are done at fixed pressure,
$P=13.5$. For the cooling protocol we always start from a Monte Carlo equilibrated configuration at
$T=5$ and quench instantaneously to $T=1$, where the system is equilibrated again using 2$\times 10^5$ Monte Carlo steps. With regular Monte Carlo we simulate three ``cooling rates". For the ``fastest" quench we take the system directly
to $T=0.1$. The next fastest protocol employs $5\times 10^4$ Monte Carlo moves at $T=1$ and then is cooled down at steps of $\Delta T=0.1$ with the same number of moves at each temperature. The next
cooling rate employs $10^5$ Monte Carlo moves after every such step. In addition to these three cooling
rates we also perform Swap Monte Carlo, starting again at $T=1$. From here we again cool at steps
of $\Delta T=0.1$ with $5\times 10^4$,  $10^5$ and $2\times 10^5$ Swap Monte Carlo moves in each step. The resulting average energy and its dependence on the inverse density is shown in Fig.~\ref{coolbin} .
To understand this figure we recall that in systems with inverse power law potential there exists
an exact relation between the average energy, pressure, density and  temperature which for Eq.~(\ref{defphi})
in two dimensions reads:
\begin{equation}
P=\rho \left(T+ 6\left\langle \frac{U}{N}\right \rangle \right) \ .
\label{relation}
\end{equation}
Inverting for the energy per particle we find in our case
\begin{equation}
\left\langle \frac{U}{N}\right \rangle  = \frac{9}{4\rho} -\frac{1}{60} \ .
\label{eqstate}
\end{equation}
We note that this relationship is not obeyed in each configuration; to stress this point we show as an example in Fig.~\ref{coolbin}
the actual values of the energy and density of randomly chosen 100 configurations out of all the configurations
that lead to the plotted average energy. Each (blue) point represents an average over 30,000 Monte Carlo steps. Indeed, the results shown in Fig.~\ref{coolbin} are in excellent agreement with the equation of state. 
This indicates that although we are always at temperature $T=0.1$ the slower rates of cooling allow
the system to compactify further to decrease the average energy per particle. The agreement with the equation 
of state indicates that the system does not fall out of equilibrium. We will see that cyclic shear is
doing exactly the same. 
\begin{figure}
	\includegraphics[width=0.36\textwidth]{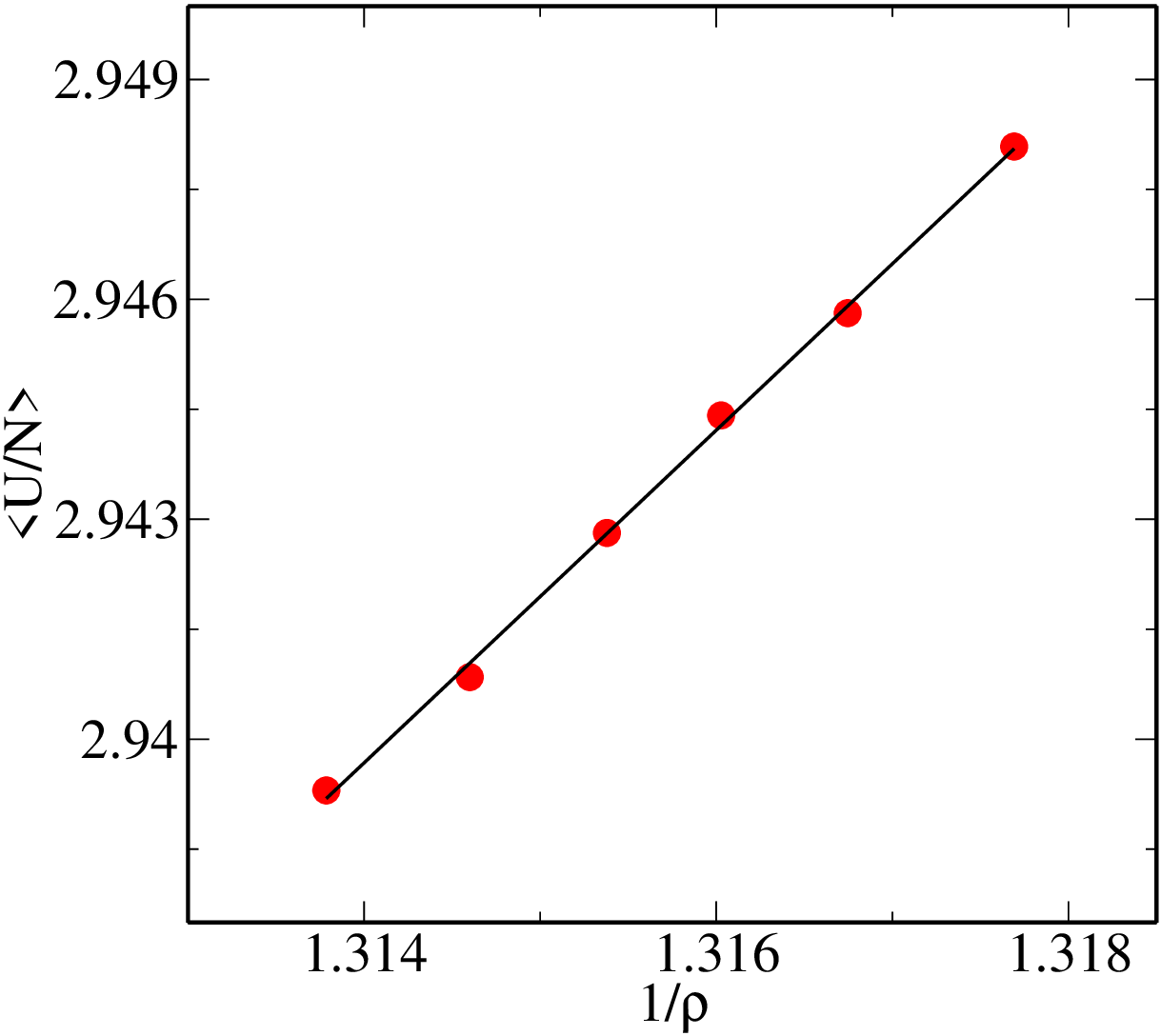}
	\caption{Average energy per particle in a binary system with $N=1024$ as a function of the inverse density
		after 1,2...6 cycles of quasi-static shear strain with $\gamma_{\rm max}=0.04$. 
		The same equation of state Eq.~(\ref{eqstate}) is obeyed}
	\label{shearbin}
\end{figure}

The cyclic shear protocol begins with instantaneous quench to $T=0.1$. After that the system is subject
to quasi-static shear strain again in $NPT$ ensemble with $P=13.5$. The strain is first increased in steps
of $\Delta\gamma=10^{-4}$ with $3000$ Monte Carlo equilibration moves after each step. Upon reaching the maximal
strain of $\gamma=0.04$ the strain is reduced with the same steps until $\gamma=-0.04$, upon which the
strain is annulled with the same steps. The thermodynamic quantities are measured at zero strain,
$\gamma=0$. The resulting average energy per particle as a function of the inverse density is shown in Fig.~\ref{shearbin}.
The linear best fit in this case has a slope of $2.27\pm 0.04$, again in very good agreement with Eq.~(\ref{eqstate}).

\begin{figure}
	\includegraphics[width=0.36\textwidth]{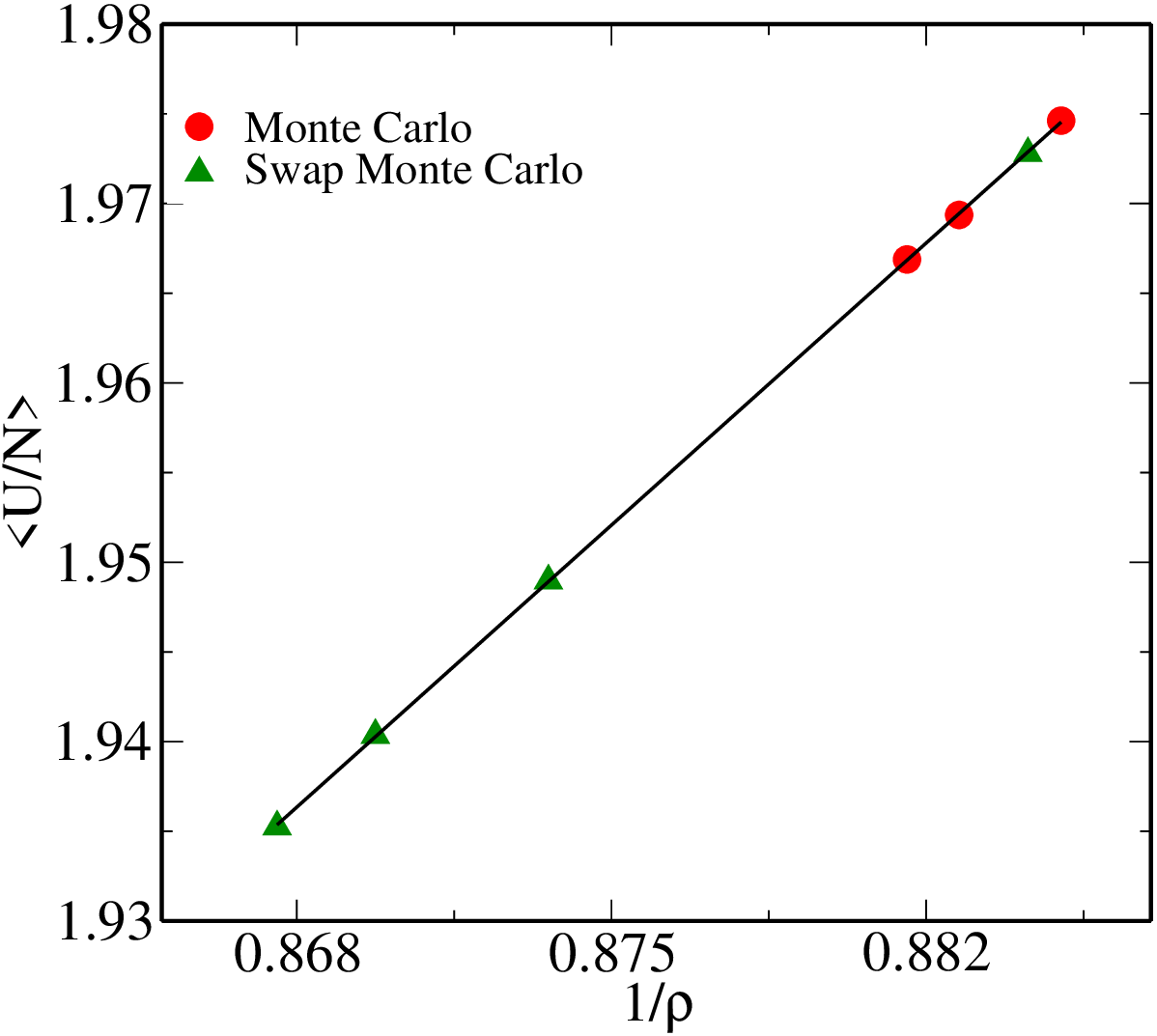}
	\caption{Average energy per particle in a poly-dispersed system with $N=1024$ as a function of the inverse density
		after cooling at seven different rates, three using regular Monte Carlo and four Swap Monte Carlo. 
		The best fit has a slope of 2.250$\pm 0.005$ in perfect agreement with the equation of state Eq.~(\ref{eqstate}).}
	\label{coolpoly}
\end{figure}

\begin{figure}
	\includegraphics[width=0.30\textwidth]{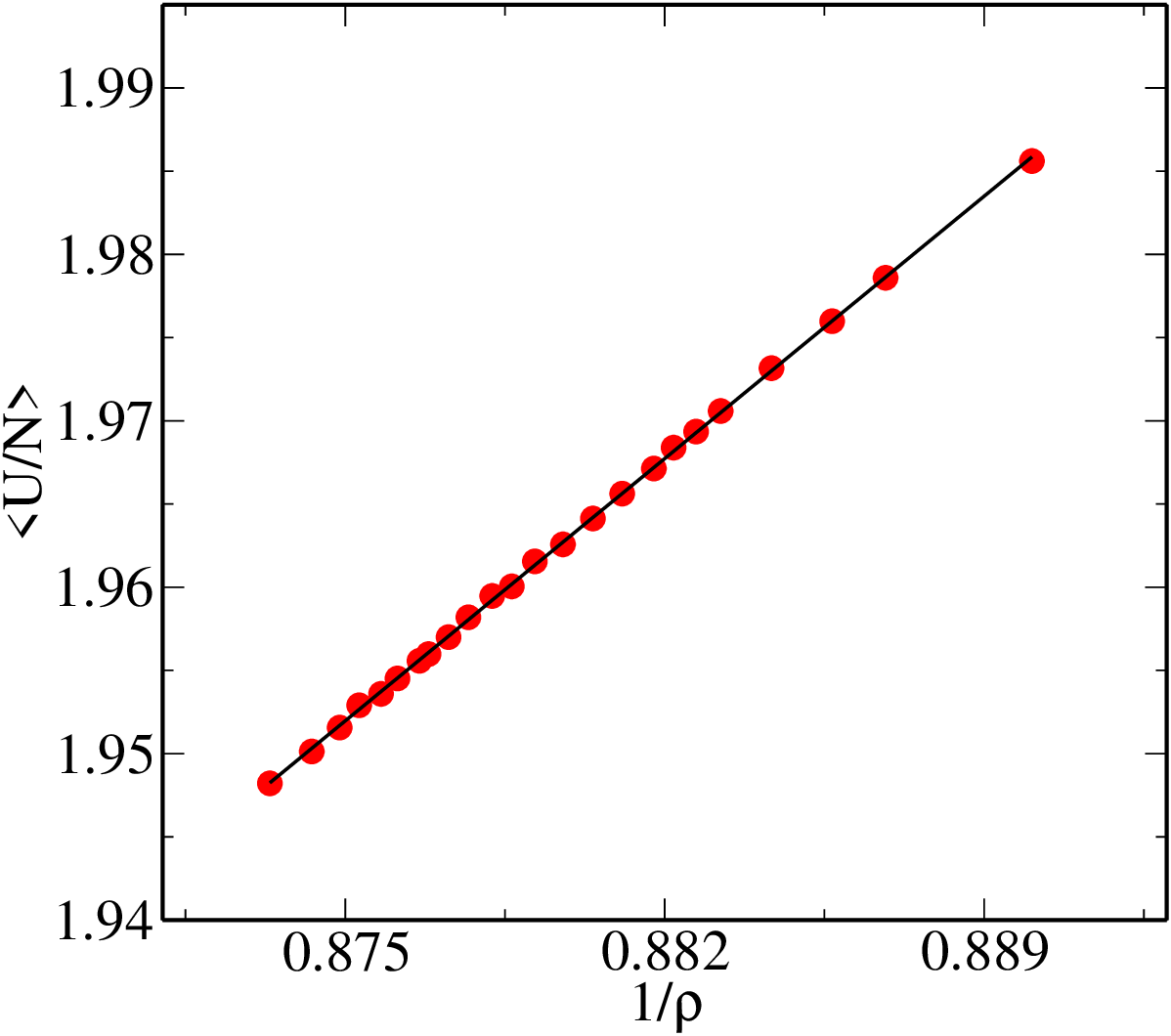}
	\vskip 0.2 cm
	\includegraphics[width=0.32\textwidth]{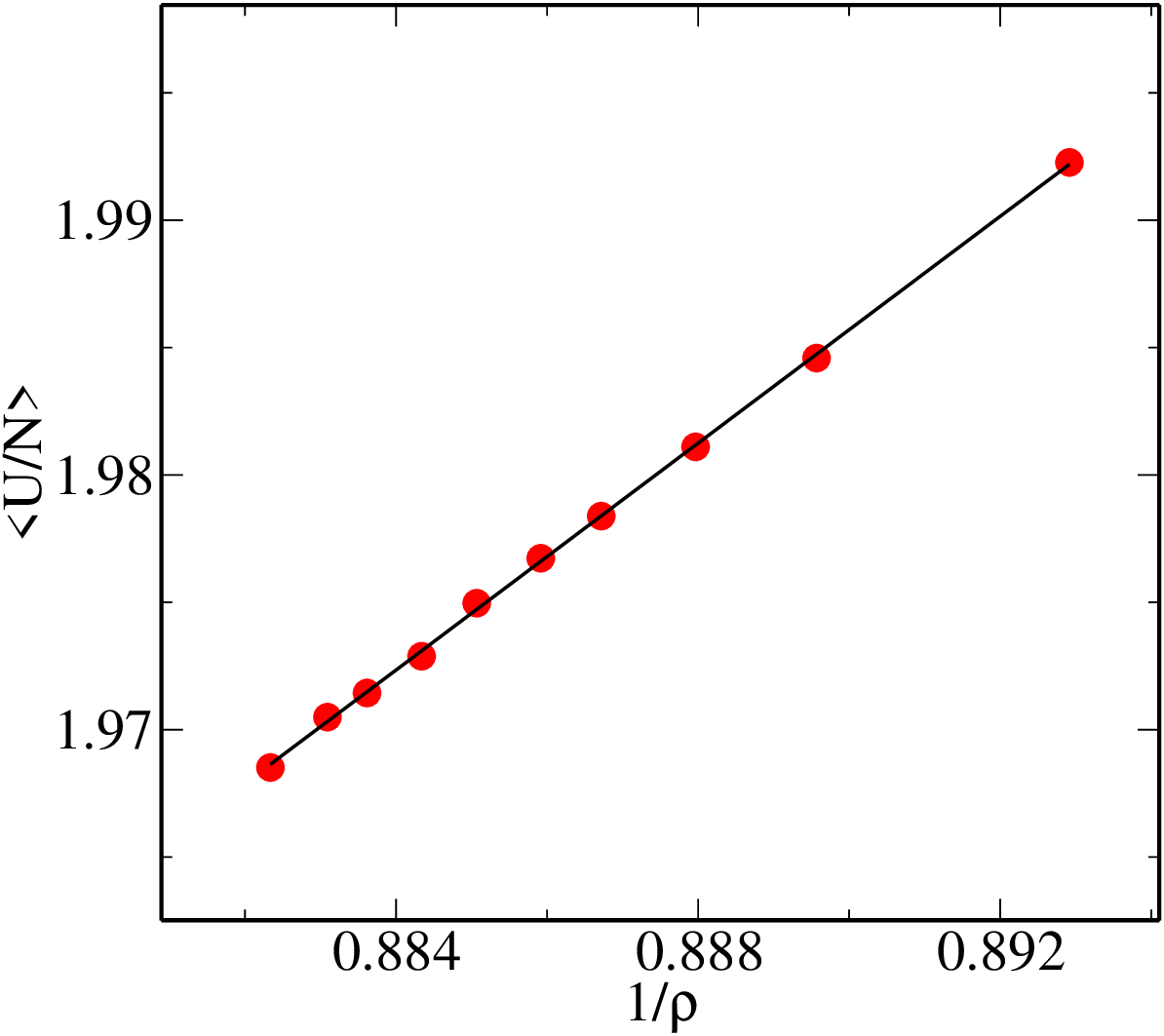}
	\caption{Average energy per particle in a poly-dispersed system with $N=400$ (upper panel) and $N=625$ (lower panel) as a function of the inverse density
		after 24 and 10 cycles respectively of quasi-static shear strain with $\gamma_{\rm max}=0.04$. 
		The same equation of state Eq.~(\ref{eqstate}) is obeyed with the slopes 2.25$\pm 0.01$  and 2.23$\pm 0.02$ respectively.}
	\label{shearpoly}
\end{figure}

Close examination of the system at the slowest cooling protocol or after six cycles of straining reveals
partial crystallization. Therefore, to ascertain that the results do not depend on this effect, we repeat
the same protocols on example II which is constructed to avoid crystallization. The very same cooling
protocol is used, but in this case we employ 4 cooling rates, with 50, 500, 5000 and $2\times 10^5$ Swap Monte Carlo steps.
Results are shown in Figs.~\ref{coolpoly} and \ref{shearpoly}. 
We find that also in the poly-dispersed example the two protocols end up with configurations where the mean energies 
agree perfectly well with the equation of state.

\begin{figure}
 	\vskip 0.2 cm
 	\includegraphics[width=0.30\textwidth]{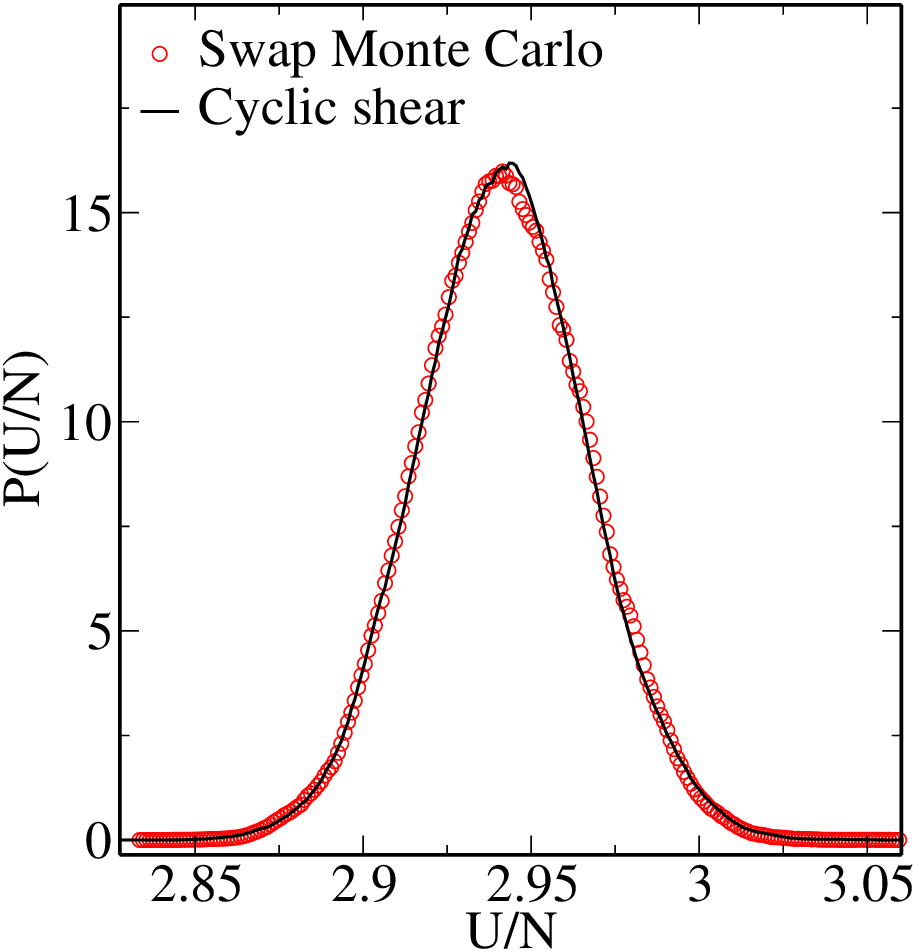}
 	\vskip 0.2 cm
 	\includegraphics[width=0.30\textwidth]{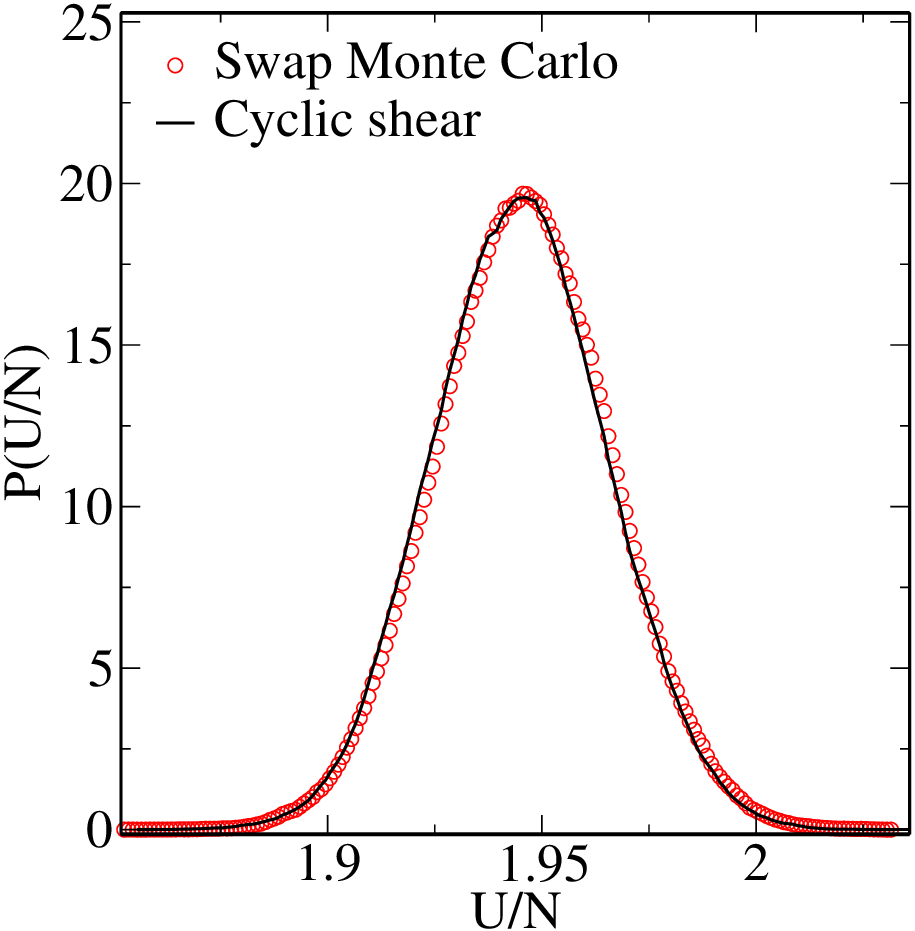}
 	\caption{Probability distribution functions (pdf's) for the total energy per particle in the binary 
 	and poly-dispersed examples (upper and lower panels respectively). In both cases the results pertain
 to $N=400$. }
 	\label{Gaussian}
 \end{figure}

Next one needs to examine the fluctuations about the average. To this aim we computed the probability distribution function (pdf) of the 
total energy $U$ of our systems, where $U=\sum_{i\ne j} \Phi(r_{ij})$. This is done simply by computing the energy of a system after each Monte Carlo step, and binning
the results as usual. The distribution are found to be Gaussian, and quite identical for the configuration found
from cooling and from cyclic shearing. An example of such Gaussian pdf's is shown in Fig.~\ref{Gaussian} for both
the binary and the poly-dispersed examples.
For this comparison we used simulation with $N=400$ for both models.

As mentioned above, a shortcoming of the vapor-deposition technique is that it results with 
anisotropic samples. One could worry that shear straining back and forth is equally likely
to produce strong anisotropies. In fact this is not the case. To see this we compute the
stress tensor $\B S$ in our simulations, defined as
\begin{equation}
S^{\alpha\beta} \equiv \frac{\sum_{i=1}^N  f^\alpha_i r^\beta_i }{L^2}
\end{equation} 
where $\B f_i$ and $\B r_i$ are the force on the $i$th particle and its position respectively.
The stress tensor has two eigenvalues $\lambda_1$ and $\lambda_2$, and the degree of
anisotropy $A$ can be estimated from \cite{19BFS}:
\begin{equation}
A\equiv \frac{\lambda_1-\lambda_2}{\lambda_1+\lambda_2} \ . 
\end{equation}
We measured $A$ in our sample after the maximal number of straining cycles at $\gamma=0$
after equilibrating with standard Monte Carlo steps. Invariably we found $A<0.01$, indicating
that anisotropy is not an issue for preparing a stable glass using this protocol.
\begin{figure}
	\includegraphics[width=0.35\textwidth]{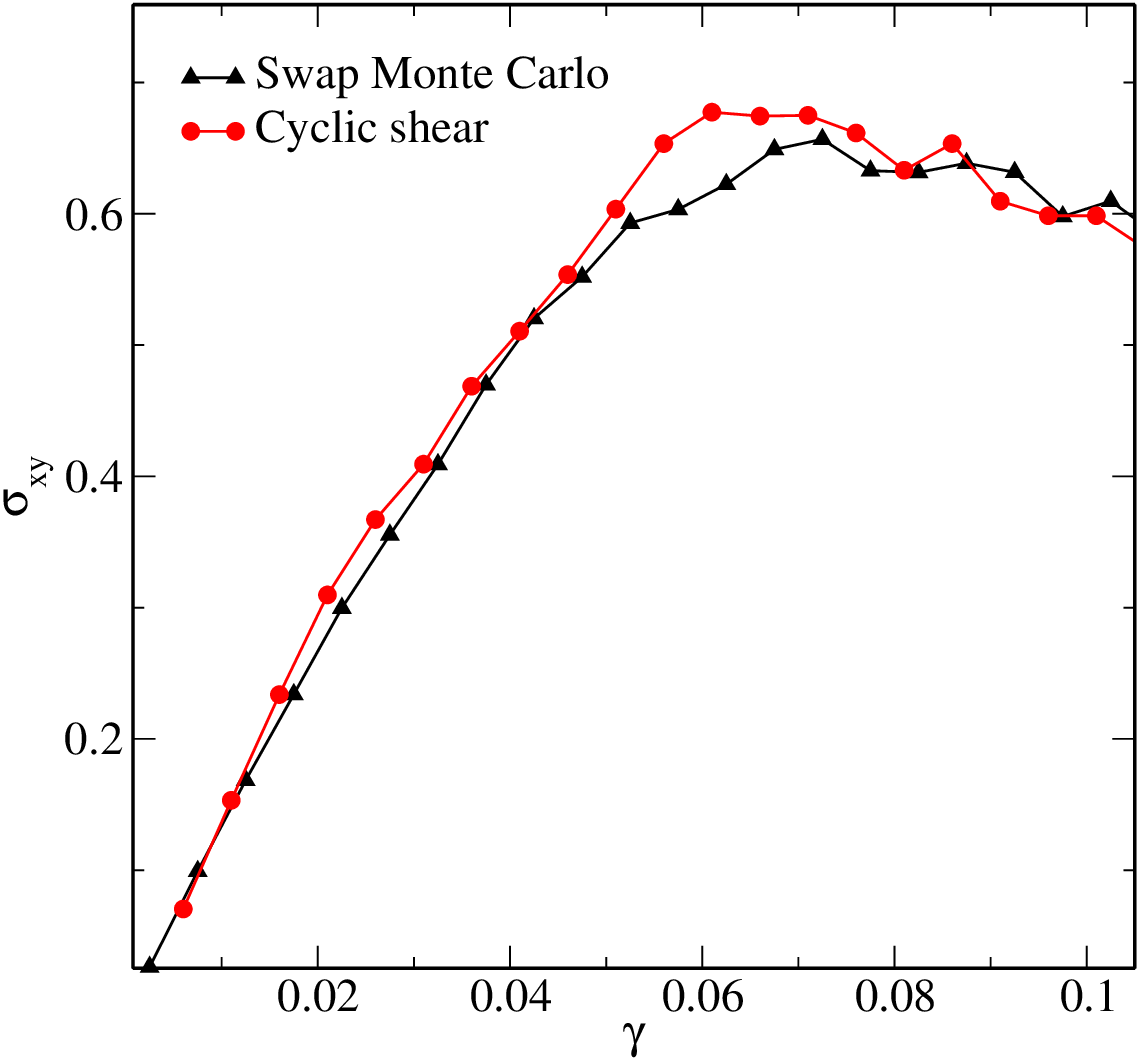}
	\caption{Athermal quasi-static stress vs. strain averaged over 30 of  poly-dispersed configuration.
		Here $N=400$ and the configuration were taken from an ensemble whose average energy is $\langle U\rangle/N=1.96$.}
	\label{svsgam}
\end{figure}

Next we compare the mechanical properties of the glasses that result in the two protocols.
To this aim we first prepared configurations either by Swap Monte Carlo or cyclic shear. Choosing
configuration from realizations contributing to close-by mean energies, we then
quenched them instantaneously to $T=0$. Finally we applied athermal quasi-static strain to measure 
the mechanical response as a function of the strain $\gamma$. Averaging over a modest number of
configurations for each protocol for $N=400$ we obtain typical results as shown in Fig. \ref{svsgam}.  
These results indicate that the mechanical response of the configurations prepared with the two
different protocols are very similar. We note that the configurations used to produce Fig. \ref{svsgam} do not 
appear particularly brittle \cite{19BFS,18OBBRT}. This is in part due to the modest size of the configurations used; each
configuration exhibits a steep drop in stress upon yield but the yield point itself is fluctuating
enough to produce a smoother average stress-strain curve. 
\begin{figure}
	\includegraphics[width=0.35\textwidth]{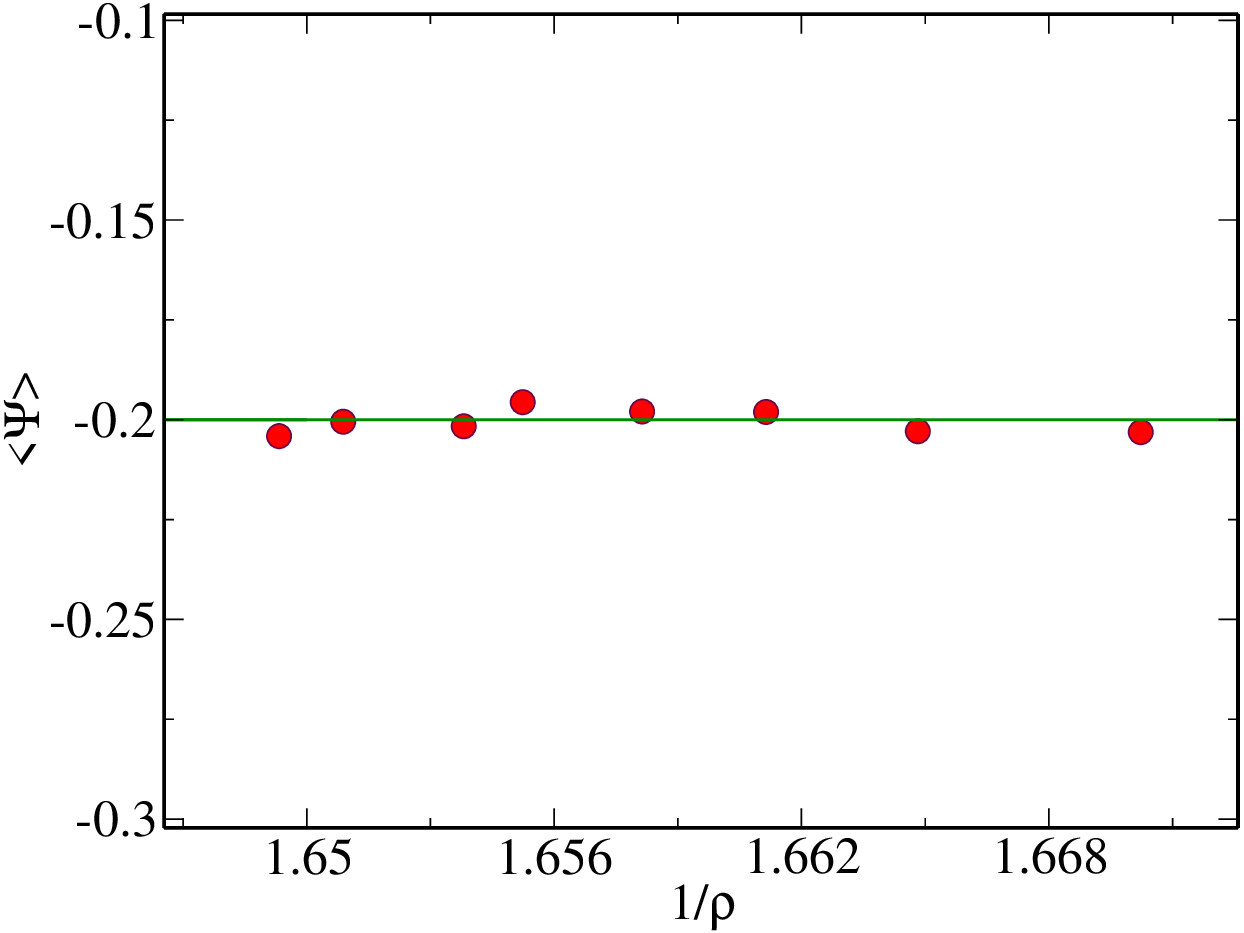}
	\caption{An example of the effect of cyclic shear on the average virial of a binary system
	with Lennard-Jones interaction at $T=0.1$. The protocol is the same as in Fig.~\ref{shearbin}. $\epsilon$ and $\sigma_{ab}$ are
	the same as in the binary systems with inverse power law potential. The 
agreement with Eq.~(\ref{simple}) is apparent, indicating that as the system compactifies it does
not fall out of equilibrium.}
	\label{generic}
\end{figure}

Finally, we need to stress that the results presented in this Letter do not depend on the potential used
being a simple inverse power law. To demonstrate this we considered a binary system with Lennard-Jones
interactions. In this generic case Eq.~\ref{relation} is changed in 2 dimensions to
\begin{equation}
P =\rho (T+\frac{\langle \Psi\rangle }{2}) \ ,
\label{relation2}
\end{equation} 
where $\Psi$ is the virial
\begin{equation}
\Psi =-\frac{1}{2N} \sum_{i\ne j,a,b} r_{ij} \frac{\partial \Phi_{ab}(r_{ij})}{\partial r_{ij}} \ .
\end{equation}
Since it has been demonstrated already that in Lennard-Jones systems Swap Monte Carlo cooling works 
well without falling out of equilibrium \cite{20POB} we will demonstrate this here only for cyclic shear.
Moreover, we will select conditions that are as far as possible from effective power-law behavior 
by choosing our NPT simulations with $P=0$ \cite{09LP}. In this case Eq.~(\ref{relation2}) simplifies
to 
\begin{equation}
\langle \Psi\rangle = -2 T \ .
\label{simple}
\end{equation}
In Fig.~\ref{generic} we show results of simulations at $T=0.1$ of the results of cyclic shear
of a 50:50 binary Lennard-Jones system. The system satisfies Eq.~\ref{simple}
quite accurately, showing that while it compactifies in never falls out of equilibrium.
Further details on this and other generic potentials will be published in a follow-up
forthcoming paper. 

In summary, we have shown that by choosing NPT conditions vs NVT ones as had been done before we could
demonstrate a quantitative similarity between configurations cooled down by Swap Monte Carlo 
and cyclic shear protocols. We compared the mean energy as a function of density, and for
a given temperature discovered that the equation of state is obeyed extremely closely. In addition, 
the fluctuations about the mean were standard Gaussian Boltzmann-Gibbs distributions that
agreed in their variance for configurations with the same mean energy. The analysis of the stress
tensor indicates that cyclic shear does not induce considerable anisotropy when $\gamma=0$. Having
thus the same statistical mechanics, one expects that also the mechanical response of the configurations
produced by the two protocols would be in agreement. This is born out in our simulations.

We should note that applying cyclic shear at finite temperature (low as it may be) and then quenching
to zero temperature, is not necessarily equivalent to quenching to zero temperature first and then
applying cyclic shear. The first protocol allows temperature fluctuations to relax the accumulation
of memory including anisotropy. This is not necessarily the case in the second scenario. Applying
loads at zero temperature can be notorious in forming memory of the whole process. 

Finally, we recall that Swap Monte Carlo protocols cannot be applied in experiments; computer simulations
allow a wider spectrum of tricks, and see for example \cite{19KJBWL}. The similarity in 
the resulting configurations discussed above provides a strong indication that cyclic shear (and probably other cyclic
loadings) can be used to achieve equilibrated well quenched amorphous solids in experiments when equilibrated cooling is not available.

We thank Edan Lerner and Srikanth Sastry for useful comments on the first draft of this Letter.
This work has been supported in part by the scientific and cooperation agreement between Italy 
and Israel through the project COMPAMP/DISORDER, the Joint Laboratory on ``Advanced and Innovative Materials" - Universita' di Roma ``La Sapienza" - WIS and the Minerva Center for "Aging, from Physical Materials to Human Tissues".  

\bibliography{ALL}

\end{document}